# Solute-Vacancy Clustering in Aluminum

Jian Peng, Sumit Bahl, Amit Shyam, J. Allen Haynes, and Dongwon Shin[*]

*Materials Science and Technology Division, Oak Ridge National Laboratory, Oak Ridge, TN 37831*

**Abstract:** We present an extensive first-principles database of solute-vacancy, homoatomic, heteroatomic solute-solute, and solute-solute-vacancy binding energies of relevant alloying elements in aluminum. We particularly focus on the systems with major alloying elements in aluminum, i.e., Cu, Mg, and Si. The computed binding energies of solute-vacancy, solute-solute pairs, and solute-solute-vacancy triplets agree with available experiments and theoretical results in literature. We consider physical factors such as solute size and formation energies of intermetallic compounds to correlate with binding energies. Systematic studies of the homoatomic solute-solute-vacancy and heteroatomic (Cu, Mg, or Si)-solute-vacancy complexes reveal the overarching effect of the vacancy in stabilizing solute-solute pairs. The binding energy database presented here elucidates the interaction between solute cluster and vacancy in aluminum, and it is expected to provide insight into the design of advanced Al alloys with tailored properties.

**Keywords:** Aluminum alloys; Solute-vacancy cluster; Binding energy; First-principles calculations; Solute-solute-vacancy binding

---

[*] Corresponding author
Email address: shind@ornl.gov (Dongwon Shin)

## 1. Introduction

Solute clusters constitute the fundamental building blocks of precipitation and age-hardening in aluminum alloys [1–7]. A few examples for various aluminum alloys follow: Cu-Mg solute clusters cause rapid hardening during the aging of Al-Cu-Mg alloys [2]. In Al-Cu alloys, a fine distribution of θ′ ($Al_2Cu$) precipitates results from heterogeneous nucleation at Sn particles that evolved from Sn clusters and, which leads to an improved age-hardening response of such alloys [3]. Mg-Si solute clusters that form during room temperature natural aging in Al-Mg-Si alloys can either decrease or increase the kinetics of precipitation depending on the aging temperature [4]. Solute cluster chemistry is reported to alter the type of precipitates that form in Al-Cu-Mg alloys [5,6]. Microalloying with Ag in Al-Cu-Mg alloys favors precipitation of the cubic Z phase over the S phase ($Al_2CuMg$) [5].

Solute atoms may combine with vacancies to form solute-vacancy and solute-solute-vacancy complexes. Interaction between solute atoms and vacancies can assist in the formation of solute clusters and precipitates through diffusion kinetics process [8–12]. The excess concentration of quenched-in vacancies available during initial stages of natural or artificial aging increases the likelihood of solute-vacancy interactions. Solute-solute-vacancy complexes, X-X-Va or X-Y-Va (X, Y are solutes and Va is a vacancy) can affect solute clustering. A recent report illustrated the annihilation of vacancies at the surface of an atom probe tomography specimen [13]. The nanoscale specimen suppressed the formation of solute clusters as opposed to the formation of solute clusters in bulk specimen of an Al-Mg-Si alloy during natural aging. In another report, vacancies injected into the microstructure by cyclic deformation produced a uniform distribution of solute clusters without precipitate free zones [14]. These observations highlight the importance of solute-vacancy interaction in the formation of solute clusters.

Vacancies in the alloy matrix not only influence the stability of atomic clusters, but also the diffusion kinetics of atomic cluster formation and precipitation pathways [15,16]. In aluminum alloys, it has been experimentally shown that small additions of Si in Al-Sc-Zr alloys accelerated the $Al_3$(Sc,Zr) precipitation kinetics [12]. First-principles calculations revealed the underlying mechanism as Si promoting the precipitation of Sc by (1) binding with Sc, (2) decreasing vacancy formation energy near Sc, and (3) lowering the Sc migration energy barrier in Al [12]. Another example is $Al_3$Sc precipitates formed as a result of interactions between Mg/Sc atoms and vacancies, which facilitated nucleation more than the pure Al-Sc binary alloy [17]. Hence, understanding the formation of solute-vacancy complexes is a critical step toward designing advanced aluminum alloys where thermodynamic stability and diffusion kinetics are manipulated to tailor the morphologies of key precipitates to achieve targeted mechanical properties [18–20].

Solute clusters are experimentally observed directly by atom probe tomography [6,7,21–25] and vacancy clusters are studied indirectly by positron annihilation spectroscopy [5,26–28]. However, fundamental insights into the formation of solute clusters are still limited and accurate determination of the binding energies of solute clusters and solute clusters with vacancies is required. It is challenging to systematically and accurately determine the various combinations of binding energies by experiments. However, first-principles calculations can predict the solute-vacancy binding energies in aluminum [29], magnesium [30], and cobalt [31] for a large number of solute elements. Also, a previous theoretical study predicted solute-solute pair binding in aluminum to explain atomistic behavior of microalloying elements during phase decompositions of Al-Cu-Mg, Al-Zn-Mg, and Al-Mg-Si alloys [32]. Although a number of experimental and theoretical studies on energetics of solute-vacancies complexes are available [10–12,18,29,32,33], a more comprehensive and consistent database that can cover various solute-vacancy clustering behavior in aluminum is required.

Herein, we present a large, comprehensive and self-consistent database of solute-vacancy pairs (X-Va), homoatomic and heteroatomic solute-solute (X-X, Cu-X, Mg-X, and Si-X) in the first nearest neighbor (1NN hereafter) and the second nearest neighbor (2NN hereafter) positions and solute-solute-vacancy triplets (X-X-Va, Cu-X-Va, Mg-X-Va, and Si-X-Va) binding energies in various positions of relevant alloying elements in an aluminum matrix from first-principles calculations based on density functional theory (DFT). The alloying elements considered in the present study are Li, Na, Mg, Si, Ca, Sc, Ti, V, Mn, Fe, Co, Ni, Cu, Zn, Ge, Zr, Nb, Ag, Cd, In, Sn, Sb, Pt and Pb [34,35]. Particular focus was paid to the systems with Cu, Mg, and Si solutes, because several Al alloy families of technological relevance are based on the Al-Cu, Al-Mg, and Al-Si systems. The overarching effect of vacancies in stabilizing solute-solute clusters is presented in this work. It is observed that solute atoms with similar chemistry based on their location in the periodic table can be grouped with respect to their clustering behavior. These groups include pre-transition metals (Li, Na, Mg and Ca), transition metals (Sc, Ti, V, Mn, Fe, Co, Ni, Cu, Zn, Zr, Nb, Ag, Cd, and Pt), post-transition metals (In, Sn, and Pb) and metalloids (Si, Ge, and Sb). A distinction is required to explain the clustering behavior within transition metals between early-transition metals (Sc, Ti, V, Zr, and Nb) and late-transition metals (Mn, Fe, Co, Ni, Cu, Zn, Ag, Cd, and Pt). Solute size and competition between Al-X and X-Y binary compound formation energies are the physical factors for solute cluster binding energies that are investigated.

## 2. Computational Method

We used 256-atoms in a 4×4×4 fcc Al supercell with defects (i.e., solutes and vacancies) to derive the binding energies of solute-vacancy pairs, solute-solute pairs and solute-solute-vacancy triplets for various pair spacing. We have exhaustively enumerated possible configurations of point defects in fcc Al supercells in 1NN and 2NN positions. We considered six different defect pairs and 23 different triplet configurations, and the schematic of defect arrangements is shown in Fig. 1.

(Fig. 1 about here)

The solute-vacancy and solute-solute pair binding energy, $E_{bind}$, is defined as follows:

$$-E_{bind}(X\text{-}X) = E(Al_{254}X_2) + E(Al_{256}) - 2E(Al_{255}X) \qquad (1)$$
$$-E_{bind}(X\text{-}Y) = E(Al_{254}XY) + E(Al_{256}) - E(Al_{255}X) - E(Al_{255}Y)$$
$$-E_{bind}(X\text{-}\square) = E(Al_{254}X\square) + E(Al_{256}) - E(Al_{255}X) - E(Al_{255}\square)$$

where X is a solute atom, Y is another solute, $\square$ is a vacancy in the Al supercell. We used positive binding energy to indicate a favorable binding to be consistent with the sign convention in literature [29,30,36]. Solute-solute-vacancy binding energy with three different reference states, i.e., with respect to (1) heteroatomic solute pair (XY) and vacancy, (2) solute (X)-vacancy pair and the other solute (Y), and (3) isolated individual defects (X, Y, and vacancy), is defined as follows:

$$-E_{bind}(X\text{-}Y\text{-}\square) = E(Al_{253}XY\square) + E(Al_{256}) - E(Al_{254}XY) - E(Al_{255}\square) \qquad (2)$$
$$-E_{bind}(X\text{-}Y\text{-}\square) = E(Al_{253}XY\square) + E(Al_{256}) - E(Al_{254}X\square) - E(Al_{255}Y)$$
$$-E_{bind}(X\text{-}Y\text{-}\square) = E(Al_{253}XY\square) + 2E(Al_{256}) - E(Al_{255}X) - E(Al_{255}Y) - E(Al_{255}\square)$$

All the DFT total energy calculations were performed with Perdew-Burke-Ernzerhof revised for solids (PBEsol) for the exchange-correlation functional [37] as implemented in the Vienna Ab initio Simulation Package (VASP) [38,39]. We have used the highly efficient graphics processing unit (GPU) version of VASP [40–42], which has proven to be as reliable as CPU-VASP for deriving interfacial and solute segregation energy [34,35] for geometrical optimization. We fully relaxed the supercells with respect to all degrees of freedom, i.e., volume, shape, and internal atom positions. We sampled k-points to achieve approximately 10,000 per reciprocal atom (4×4×4 Monkhorst-Pack grid) and set the energy cutoff to 520eV. We have performed spin-polarized calculations with the interpolation scheme suggested by Vosko et al. [43] as implemented in VASP for magnetic solute elements, i.e., Cr, Mn, Fe, Co, and Ni. We also have considered spin-polarized calculations for V [30]. The accuracy of the present study may be systematically improved by including the vibrational contributions in aluminum, as shown in the case of Al-Si and Al-Cu binary systems [44,45]. However, it has been shown that formation entropy of point defects, such as vacancy and anti-sites, is much smaller in magnitude than those of intermetallics [46]. Hence, we have not considered vibrational contribution in the present work. We have computed a total of 1,621 supercells: 24 single solutes, 260 pair defects, and 1,377 solute-solute-vacancy triplets. The binding energy in Eqn. (2) was determined from the lowest total energy of the supercell of a given defect cluster.

## 3. Results

### 3.1. Solute-vacancy clustering

Fig. 2a shows the DFT binding energies in 1NN and 2NN for the X-Va pair. The corresponding numerical data are reported in Supplementary Table S1. We compared our predicted solute-vacancy binding energies of select elements with previous theoretical studies and experiments in literature, and the comparisons are summarized in Table 1. Most of our results are in good agreement with previously reported values, although there is a discrepancy in solute-vacancy binding energies between unrelaxed [33] and fully relaxed supercells [29] in Cu-Va. Gorbatov et al. [18] also compared the difference between relaxed and unrelaxed supercells. They found a peculiar behavior involving Cu solutes that the Cu-Va pair in an unrelaxed supercell is repulsive (~-0.05eV) while it is weakly favored (0.015eV) in a relaxed structure. In certain supercells of our fully relaxed calculations, there was a slight distortion in the cubic structure, but most of Al atoms maintained the equilibrium bonding distance of fcc Al.

(Fig. 2 about here)

(Table 1 about here)

For most of the solutes considered in this work, their binding energies with a vacancy in 1NN are positive and larger than in 2NN, indicating that they prefer to cluster with a vacancy in 1NN rather than in 2NN. Li, early transition metals (Nb, Ti, Zr, V, and Sc), and late transition metal (Mn) have negative binding energies with a vacancy in 1NN but positive binding energies in 2NN, revealing that clustering of these solutes with a vacancy is favorable in 2NN. The maximum binding energy between 1NN and 2NN for the X-Va pair is plotted in Fig. 2b. The expected cluster configuration in Al alloys will be determined based on the maximum value of binding energy between 1NN and 2NN. All the maximum binding energies are positive among the elements considered in the present work, indicating these elements prefer to bind with a vacancy in either 1NN or 2NN. Post transition metals (In, Sn, and Pb) are a consistent group of elements with one of the most favorable X-Va binding energies.

## 3.2. Homoatomic solute-solute pair binding

Binding energies of homoatomic solute-solute pairs in 1NN and 2NN positions are shown in Fig. 3a. Hirosawa et al. [32] also have computed the same properties and their values are overall in good agreement with ours except transition metal pairs, i.e., V, Mn, Fe, Co, Ni and Cu (see Supplementary Figure S1). We believe this discrepancy is attributed to the different relaxation scheme, as discussed by Wolverton [29]. As shown in Fig. 3a, the DFT binding energies in 1NN and 2NN for the X-X pairs can be divided into three categories. The region in the upper left corner has a positive binding energy in 2NN position and the 2NN binding energy is always larger than in 1NN, indicating that pre-transition metals (Li, Mg, and Ca), post-transition metal (Sn) as well as early transition metals (Sc, Ti, Nb, and Zr), prefer to form a X-X cluster in 2NN position rather than in 1NN. It is intriguing that most of these elements (Li, Sc, Ti, and Zr) are known to form $L1_2$-trialumindes, whose atomic position is in the distance of 2NN. Elements located in the bottom right trapezoid have positive binding energy in 1NN, and the binding energy in 1NN for these elements is always larger than in 2NN, meaning these elements prefer to form X-X cluster in 1NN. These elements are pre-transition metal (Na), late transition metals (Fe, Co, Ni, Cu, Zn, Ag, Cd, and Pt) and post-transition metals (In and Pb) and metalloids (Ge and Sb). Particular attention should be paid to solutes V, Mn and Si, which have negative binding energies in both 1NN and 2NN, indicating it is energetically not favorable for these elements to form X-X cluster in both 1NN and 2NN. Fig. 3b shows the maximum binding energy between 1NN and 2NN of each element. No clear trend within each group is observed for X-X clusters.

(Fig. 3 about here)

3.3. Heteroatomic (Cu, Mg, Si)-solute (X) pair binding

The behavior of the considered elements varies significantly when binding with Cu, Mg, or Si in the Al matrix. Fig. 4 shows the DFT binding energies in 1NN and 2NN for Cu-X, Mg-X and Si-X clusters. In the Cu-X system (Fig. 4a), only Si prefers to bind with Cu in 2NN. Early transition metals (V, Nb, Ti, Sc, and Zr) are repulsive to Cu in both 1NN and 2NN, indicating forming a cluster between these elements and Cu is not favorable. For remaining elements, forming a Cu-X cluster in 1NN is preferred rather than in 2NN. For the Mg-X system (Fig. 4b), pre-transition metals (Li, Na, Mg and Ca) and early transition metals (Sc, Ti, Zr and Nb) prefer to bind with Mg in 2NN rather than in 1NN. Elements such as late transition metals (Ni, Cu, Zn, Ag, Cd and Pt), metalloids, and post-transition metals have a favorable binding with Mg in 1NN. Late transition elements (Mn, Fe, Co and V) have negative binding energies in both 1NN and 2NN positions. In contrast, in the Si-X system (Fig. 4c), pre-transition metals (Li, Na, Mg, and Ca) and early-transition metals (Sc, Ti, Zr and V) prefer to bind with Si in 1NN rather than in 2NN. Late-transition metals (Fe, Co, Ni, Cu, Zn, Ag, Cd and Pt) prefer to bind with Si in 2NN. Fig. 4c also reveals that it is not favorable for Si to form a Si-X cluster with metalloids Ge, Si and Sb, and post-transition metals In and Sn. Even the post-transition metal Pb has a weakly positive 1NN binding energy with Si.

(Fig. 4 about here)

(Fig. 5 about here)

The maximum binding energies of the Cu-X, Mg-X and Si-X clusters in 1NN and 2NN are presented in Fig. 5. Late-transition metals (Mn, Fe, Co, Ni, Cu, Zn, Ag and Pt) have positive Cu-X binding energies, whereas early-transition metals (Sc, Ti, V, Zr and Nb) have negative Cu-X binding energies. Overall, late-transition metals are a consistent group of elements with one of the most favorable Cu-X binding energies. The trend in binding energy reverses significantly in Mg-X, compared to Cu-X. Early-transitions elements (Sc, Ti, Zr and Nb) with negative Cu-X binding energies have positive Mg-X binding energies. Similarly, late-transitions metals (Mn, Fe and Co) with positive Cu-X binding energies have negative Mg-X binding energies. In further contrast with Cu-X, post-transition metals (In, Sn, and Pb) are a consistent group of elements with one of the most favorable Mg-X binding energies. The trend in maximum binding energies in the Si-X system is different from those in both Cu-X and Mg-X clusters. None of the early and late transition metals have negative Si-X binding energies. The Si-X binding energy decreases monotonically with an increase in atomic number within groups of 3d (Sc, Ti, V, Mn, Fe, Co, Ni, Cu and Zn) and 4d transition metals (Zr, Nb, Ag and Cd). Post-transition metals and metalloids with positive Cu-X and Mg-X binding energies have either negative or weakly positive Si-X binding energy.

3.4. X-X-Va, Cu-X-Va, Mg-X-Va and Si-X-Va triplets

Fig. 6 compares the maximum binding energy of binary and ternary pairs in X-X-Va, Cu-X-Va, Mg-X-Va and Si-X-Va systems. The corresponding numerical data are reported in Supplementary Table S2. In most cases, Fig. 6a reveals that the maximum binding energy of the X-Va cluster is higher than its homoatomic X-X counterpart, indicating that most solutes prefer to bind with Va. Transition metals are the exceptions. Transition metals Sc, Ti, Fe, Co, Ni, Cu and Zr prefer to form homoatomic X-X pair rather than X-Va pair. This phenomenon is also applicable to heteroatomic clusters of Mn, Fe, Co, and Ni in the Cu-system (Fig. 6b), and Ti, V, Mn, Fe, Zr and Nb in the Si-X system (Fig. 6d). However, this case is not true in the Mg-X system. Except for Nb in the Cu-X-Va system, we found that the triplets (X-X-Va, Cu-X-Va, Mg-X-Va, and Si-X-Va) always have the highest binding energy compared with their binary counterparts, revealing the triplets are the most stable clusters and a vacancy can stabilize the binary solute-solute cluster.

(Fig. 6 about here)

## 4. Discussion

4.1. Physical contributions to solute-solute bindings in Al

*4.1.1. Solute size*

A moderate correlation between solute size and solute-vacancy binding energies has been reported [29], with the larger the solute, the higher the binding energy. This trend was explained by a simple strain model [29]; placing a vacancy next to a large solute atom in the Al matrix allows the solute atom to relax towards the vacancy. Consequently, being away from the other neighboring Al atoms helps to relieve the strain induced by the large solute and produces stronger solute-vacancy binding. Fig. 7 relates the atomic size of solute X with the X-X, X-Va, Cu-X, Mg-X and Si-X binding energies. Across these binary systems, a modest positive correlation between the solute size and the 1NN binding energy was also observed: solutes in the same column of the periodic table, Cu/Ag, Zn/Cd, Si/Ge/Sn/Pb in all binary systems and Mg/Ca in systems except X-X, and solutes in the same row of the periodic table, Co/Ni/Cu/Zn in systems except X-Va, Ag/Cd/In in the X-X and Cu-X systems, and Ti/V/Cr/Mn/Fe/Co/Ni/Cu/Zn/Ga/Ge in the Si-X system.

Since both Cu and Si atoms are smaller than the Al atom, the simple strain model is still applicable to the Cu-X and Si-X systems. Even though this model is invalid for the Mg-X system since Mg is larger than Al, a moderate and positive correlation between solute size and binding energy is still observed. In addition to that, the inability of atomic size to explain solute-vacancy binding energies was particularly noted for 3d transition metals [29]. Our results also highlight the inability of atomic size alone to explain the solute-solute binding energies for all of the transitions metals. This observation reveals that solute size is a contributing factor to the solute-vacancy and solute-solute binding energy, but not the key determinant. There are other factors that govern the solute-solute binding energy.

(Fig. 7 about here)

*4.1.2. Lowest formation energies of binary intermetallic compounds*

The affinity between solute and matrix atoms has been used to understand the solute-vacancy binding energies, particularly for 3d transition metals where atomic size does not correlate with the solute-vacancy binding energies [29]. A low solute-vacancy binding energy is expected when a strong bond exists between matrix and solute atoms. Strong matrix-solute bonding makes it energetically unfavorable for a solute to bind to a vacancy. The interatomic bonding argument was presented elsewhere [29], but was primarily qualitative. One quantitative measure of affinity between two different atoms is the formation energy of a binary intermetallic compound comprising those atoms. We compared the solute-solute (Cu-X, Mg-X and Si-X) and Al-solute (Al-X) formation energies of binary intermetallics for all the elements investigated. Although it is simple, this energetic comparison can be used to examine the correlation between 1NN and 2NN configuration and the maximum binding energy for solute-solute clusters.

(Fig. 8 about here)

Fig. 8 compares the lowest formation energies of the compounds in the Cu-X, Mg-X, and Si-X binary systems with those in the Al-X binary system. All these formation energies were obtained from the open quantum materials database (OQMD) [47]. The diagonal line in each graph represents equal formation energies for Al-X and Cu-X, Mg-X or Si-X systems. Ideally, it is expected that solute atoms located above the lines will prefer to form Cu-X, Mg-X and Si-X clusters in the 1NN configuration. Solute atoms located below the lines will either form Cu-X, Mg-X or Si-X clusters in the 2NN configuration or will not form these clusters. In the case of Cu-X clusters, Fig. 8a agrees with our hypothesis that clusters with early transition metals (Sc, Ti, V, Zr and Nb) do not form in the 1NN configuration. However, it is in disagreement with our prediction that Cu-X clusters of several late-transition metals (Fe, Co, Ni, Ag and Pt) do not form in the 1NN configuration. Fig. 8a also finds the same correlation of the favorable configuration of Cu-X clusters of post-transition metals, but fails for most of the pre-transition metals and metalloids. For the Mg-X cluster, Fig. 8b is able to identify the configuration for all early-transition metals, post-transition metals and metalloids. The configuration of the majority of Mg-X clusters with late-transition metals (excluding Ni, Cu and Pt) and pre-transition metals (excluding Na) is also correctly found. As shown in Fig. 8c, the configurations in Si-X clusters are correctly projected for all early-transition metals (except Nb), pre-transition metals, post-transition metals (except Pb) and metalloids. The configuration of the majority of Si-X clusters with late-transition metals (excluding Mn and Fe) is also correctly predicted. Thus, the formation energy of binary intermetallic compounds can be used as an indicator of whether solute clusters will form in 1NN configuration for Mg-X and Si-X clusters and to a lesser extent for Cu-X clusters.

(Fig. 9 about here)

The effectiveness of lowest compound formation energies in explaining the trend in the values of binding energies, apart from explaining the solute cluster configuration, was further investigated. Fig. 9 shows the relationship between maximum binding energy and lowest compound formation energy for elements in Cu-X, Mg-X and Si-X clusters. The solute cluster configuration pertaining to the maximum binding energy is indicated by separate colors. The elements with maximum binding energy in the 1NN configuration are colored in black while those in 2NN configuration or with negative maximum binding energies are colored in red. Ideally, a more negative compound formation energy should imply higher binding energy. However, the trend observed in the Cu-X clusters contradicts this ideal trend that elements with more negative Cu-X formation energy have lower Cu-X binding energy, irrespective of 1NN or 2NN configurations. The relationship between maximum binding energies of Mg-X clusters and Mg-X compound formation energy depends on the cluster configuration. The ideal trend of higher binding energy at more negative compound formation energies is followed only for elements that form clusters in 1NN configuration. Thus, Mg-X compound formation energies can explain the binding energies of only those Mg-X clusters that form in 1NN configuration. Similarly, compound formation energies of Si-X can explain the trend in binding energies of only those Si-X clusters that form in 1NN configuration.

(Fig. 10 about here)

4.2. Mechanistic insights into solute clustering through solute-solute-vacancy binding energies

Solute clusters can have a dramatic impact on the precipitation behavior and consequently, the mechanical and thermal properties of an alloy. The trends of solute-solute-vacancy binding energies emphasize the overarching effect of vacancies on enhancing the stability of solute clusters in the Al matrix as shown in Fig. 10. Intriguingly, L1$_2$-trialuminide forming elements (Li, Zr, Ti, and Sc) do not have a notable effect of stabilizing homoatomic pairs and all these elements have almost identical binding energies with and without vacancies (see Fig. 10a). Another exception is in the Cu-Nb-Va cluster among all the analyzed clusters (see Nb below the diagonal line in Fig. 10b).

The stabilization effect of vacancies can be rationalized with several experimental observations as described below. Experimental observations are chosen based on the roles of solute clusters in Al alloys. Examples from alloy systems are intended to be illustrative and not exhaustive. We expect other researchers to obtain additional mechanistic insights based on the presented database. In Al alloys, solute clusters can act as sites for heterogeneous nucleation of precipitates [3,48–52], accelerate or decelerate precipitation kinetics [4,48,53–55], select the type of precipitates that form [5,56,57] and, be strengtheners themselves [2,58–61]. Some examples that do not belong to any of these four categories, but underscore the importance of vacancies in stabilizing solute clusters, are also discussed.

*4.2.1. Al-Cu-Sn alloy*
Solute clusters can transform into particles at a particular stage of aging, which has been observed in a Sn modified Al-Cu alloy. The alloy in the as-quenched state formed Sn-Sn clusters, which eventually transformed into much bigger Sn particles during the early stages of artificial aging [3]. Direct nucleation of θ′ occurred on Sn particles without the formation of intermediate GP zones and θ″ precipitates. The fine precipitation of θ′ resulted in an enhanced age-hardening response of the Al-Cu alloy with Sn. The extremely large Sn-Sn-Va binding energy (0.48eV) identified in the present work supports their observation of pure Sn clusters. The suppression of GP zones and θ″ phases implied a shortage of free vacancies in the matrix necessary for their precipitation [3]. This observation supports the proposition that vacancies were instead sequestered inside Sn clusters. Pure Cu clusters were also observed albeit infrequently as expected based on the significantly lower Cu-Cu-Va binding energy (0.10eV) compared to Sn-Sn-Va binding energy (0.48eV) as shown in Fig.10a.

*4.2.2. Al-Mg-Si alloy*

Solute clusters can accelerate or decelerate aging kinetics depending upon thermal history and aging temperature and, consequently, affect the age hardening behavior. It is reported that the Mg-Si clusters that form in the Al-Mg-Si alloy during long-term natural aging can decelerate the aging kinetics at lower artificial aging temperatures and accelerate the aging kinetics at higher artificial aging temperatures, compared with the alloy aged directly after quenching [4]. The Mg-Si clusters sequester vacancies during natural aging, thus preventing their annihilation. The clusters release the sequestered vacancies at higher aging temperatures to accelerate aging kinetics but retard aging kinetics at lower aging temperatures due to continued sequestration. Addition of Sn to Al-Mg-Si further enhances the aging kinetics of naturally aged alloys at higher aging temperatures [62]. Present calculations indicate that the Sn-Sn-Va cluster has a very large binding energy (0.48eV) compared with that of Mg-Si-Va (0.11eV). Consequently, Sn can sequester more vacancies, resulting in an enhancement of the aging kinetics during high-temperature artificial aging. No hardening occurs during natural aging, implying the absence of the Mg-Si clusters, which is probably again due to a lack of free vacancies.

APT observations have confirmed the presence of homoatomic Si-Si clusters in the as-quenched state as well as after natural aging at room temperature for one week in an Al-Mg-Si based alloy [63]. The observations contradict the slightly negative binding energy of Si-Si cluster (-0.01eV), i.e., unfavorable binding of Si with Si (see Fig. 10a). Nevertheless, the present results show the Si-Si clusters can be stabilized by vacancies, indicated by the positive binding energy of the Si-Si-Va cluster (0.14eV). Gorvatov et al. [18] also predicted that vacancies can stabilize the formation of Si-Si clusters in Al. Therefore, we postulate that the Si-Si cluster observed in the Al-Mg-Si alloys [63] may actually be a Si-Si-Va complex.

4.2.3.  Al-Cu-Mg alloy

Modification in the chemistry of solute clusters by additional alloying elements can affect the type of precipitate that is formed. For example, microalloying an Al-Cu-Mg alloy with Ag suppressed the formation of the orthorhombic S phase and favored the precipitation of the cubic Z phase [5]. The Mg:Cu ratio is higher in the Z phase (close to two compared to one in the S phase). It was observed that the concentration of Mg increased in the solute cluster-vacancy complex due to the addition of Ag [5]. The higher concentration of Mg in the solute cluster-vacancy complex due to the presence of Ag may have influenced the preference of the Z phase over the S phase in this alloy. The Mg-Ag-Va cluster has the highest binding energy (0.19eV) compared with other possible ternary clusters, such as Ag-Ag-Va (0.16eV), Mg-Mg-Va (0.06eV), Cu-Cu-Va (0.10eV), and Cu-Ag-Va (0.15eV) as shown in Fig. 10. This trend in calculated binding energy values supports the experimental observation that Ag addition increases the concentration of Mg in the solute-vacancy complexes [5].

Al-Cu-Mg alloys are known to undergo rapid hardening within the first few tens of seconds of artificial aging at elevated temperatures (110 – 240 °C) [2]. These alloys can reach up to 70% of their maximum hardness during this stage [2]. The rapid hardening is attributed to the Cu-Mg heteroatomic clusters, while Cu-Cu and Mg-Mg homoatomic clusters are less important in hardening. The binding energy of the Cu-Mg cluster (0.024eV) is comparable to the Mg-Mg cluster (0.023eV) but significantly lower than the Cu-Cu cluster (0.06eV) (see Figs. 10a and 10b). However, this trend is inconsistent with the experimental observation that number density of the Cu-Mg clusters is comparable to Cu-Cu or Mg-Mg clusters and a few Cu-Mg cluster even grow the largest among all in size [2]. This finding can be justified if Cu-Cu-Va, Mg-Mg-Va and Cu-Mg-Va binding energies are considered. The binding of Cu-Mg is significantly favored by vacancy by having triplet binding energy of 0.09eV from pair binding of 0.02eV (0.07eV in difference). This magnitude is much higher than those of the Cu-Cu (0.096eV from 0.06eV; 0.036eV in difference) and Mg-Mg clusters (0.06eV from 0.023eV; 0.037eV in difference). Thus, it can be concluded that vacancies are critical in stabilizing the Cu-Mg cluster, which is responsible for the rapid hardening phenomenon.

*4.2.4. Al-Zn-Mg alloy*

Precipitate-free zones form near grain boundaries during aging due to annihilation of vacancies at the grain boundaries in precipitation hardened systems [64]. The importance of vacancies in stabilizing solute clusters can be gauged by analyzing solute cluster formation near grain boundaries as was done in the case of an aged Al-Zn-Mg alloy [21]. The number density of Mg-Zn clusters was lower in the region of precipitate free zone closer to the grain boundary. Vacancy depletion is enhanced in the region closer to the grain boundary, where the number density of Mg-Zn clusters reduced [21]. Although vacancies promote cluster formation, solute concentration in the matrix generally far exceeds the concentration of vacancies. Solute clusters are still expected to form in this case if their binding energies are positive or additional driving forces such as external mechanical strains are provided. In a recent report, precipitate free zones were found to be absent in Al alloys cyclically deformed at room temperature [14]. A spatially uniform distribution of vacancies was produced by cyclic deformation, which prevented the formation of precipitate free zones.

## 5. Conclusion

We report a large DFT database of solute-vacancy clustering in 1NN and 2NN positions of relevant alloying elements in aluminum. We considered the defect configurations of homoatomic and heteroatomic solute-solute, and solute-solute-vacancy binding. Particular focus is given to heteroatomic solute-solute Cu-X, Mg-X, and Si-X binding energies due to the technological importance of these alloying elements in aluminum alloys. The binding energies are observed to have different variations with solute X in Cu-X, Mg-X, and Si-X systems.

Physical factors such as solute size and formation energies of corresponding binary compounds correlated with the solute-solute binding energies. A moderate and positive correlation between the solute size and solute-vacancy and solute-solute binding energies was observed. We found a correlation between formation energies of Cu-X, Mg-X, and Si-X with Al-X binary intermetallic compounds and the solute-solute binding behavior to a greater extent for Mg-X and Si-X systems and to a lesser extent for Cu-X system. The variation in binding energy can be correlated with those solute-solute clusters that have binding energy higher in 1NN than 2NN configuration. Essentially, a more negative (favorable) solute-solute compound formation energy compared to Al-solute leads to 1NN configuration and higher solute-solute binding energy.

Vacancies are found to have an overarching effect on stabilizing solute-solute clusters with a few exceptions: homoatomic pairs of Li, Sc, Ti, and Zr (known to form $L1_2$-trialuminide) and Cu-Nb pair. The solute-solute-vacancy triplet binding energies are demonstrated to explain several experimental observations in literature. Solute-solute-vacancy complexes can have substantial effects on the precipitation behavior and, thus, the mechanical and some functional properties of Al alloys. The large binding energy database presented here enhances our understanding of the solute cluster and vacancy interaction for Al alloys and it is anticipated to provide insight into the design of new Al alloys with tailored properties.

**Acknowledgement**

Research was supported by the U. S. Department of Energy, Office of Energy Efficiency and Renewable Energy, Vehicle Technologies Office, Propulsion Materials Program. Early research was sponsored by the Laboratory Directed Research and Development Program of Oak Ridge National Laboratory, managed by UT-Battelle, LLC, for the U. S. Department of Energy. This research used resources of the Oak Ridge Leadership Computing Facility at the Oak Ridge National Laboratory, which is supported by the Office of Science of the U. S. Department of Energy under Contract No. DE-AC05-00OR22725.

**List of Tables**

Table 1 Comparison between calculated and experimental values of solute-vacancy binding energies of 1NN position (eV) in aluminum. Positive numbers represent favorable binding between solute and vacancy.

Table 1 Comparison between calculated and experimental values of solute-vacancy binding energies of 1NN position (eV) in aluminum. Positive numbers represent favorable binding between solute and vacancy.

| Solute-vacancy pair | Present study[a] | DFT[b] | DFT[c] | DFT[d] | DFT[e] | Exp.[g] |
|---|---|---|---|---|---|---|
| Mg-□ | 0.01 | -0.02 | 0.01 | 0.02 | 0.011 | -0.01±0.04 |
| Si-□ | 0.06 | 0.08 | 0.05 | 0.03 | 0.051 | 0.03 |
| Sc-□ | -0.12 | -0.17 | -0.12 | -0.12 | | |
| Cu-□ | 0.03 | ~0.02 | ~0 | -0.05 | 0.015 ~-0.05[f] | 0.00±0.12 |
| Zn-□ | 0.05 | 0.03 | 0.04 | 0.03 | 0.035 | 0.02 |
| Zr-□ | -0.21 | -0.28 | -0.20 | -0.15 | | |
| Ag-□ | 0.11 | 0.07 | 0.09 | 0.09 | | |

[a]DFT, PBESol
[b]Ref. [29], local density approximation (Mg, Zr, Sc, Zn and Ag), generalized gradient approximation (Si, Cu)
[c]Ref. [10], Perdew-Burke-Ernzerhof
[d]Ref. [32], Korringa-Kohn-Rostoker Green's function
[e]Ref. [18], cluster expansion/projector augmented wave
[f]Unrelaxed
[g]Ref. [11]

**List of Figures**

Figure 1 Schematics of atomic arrangements of solute-vacancy, solute-solute pairs and solute-solute-vacancy triplets within the Al supercell. Only the corners of respective 256-atoms fcc Al supercell (4×4×4) are shown.

Figure 2 (a) DFT binding energies in 1NN and 2NN for the X-Va pair and (b) maximum values of X-Va binding energies grouped according to the location of X in the periodic table. Within each group, elements are arranged in an increasing order of atomic number. The numerical data is reported in Supplementary Table S1.

Figure 3 (a) DFT binding energies in 1NN and 2NN for the X-X pair, and (b) Maximum values between the 1NN and 2 NN binding energies according to the location of solute X in the periodic table. Within each group, the elements are arranged in an increasing order of atomic number. The numerical data are reported in Supplementary Table S1.

Figure 4 DFT binding energies in 1NN and 2NN for the (a) Cu-X, (b) Mg-X and (c) Si-X pairs. The numerical data are reported in Supplementary Table S1.

Figure 5 Maximum values between the 1NN and 2 NN binding energies of the Cu-X, Mg-X and Si-X systems grouped according to location of X in the periodical table. The numerical data are reported in Supplementary Table S1.

Figure 6 Maximum binding energies (in eV) of various possible grouping in (a) X-X-Va (b) Cu-X-Va, (c) Mg-X-Va and (d) Si-X-Va clusters. The numerical data are reported in Supplementary Table S2.

Figure 7 Solute size and DFT 1NN binding energies of the X-X, X-Va, Cu-X, Mg-X, and Si-X clusters.

Figure 8 Comparison of the lowest formation energies of the compounds in the (a) Cu-X, (b) Mg-X and (c) Si-X systems with their counterparts in the Al-X system. These data were collected from the open quantum materials database (OQMD) [32]. The black line indicates where the lowest formation enthalpy of the Cu-X, Mg-X or Si-X compounds are equal to that of the Al-X compounds. The elements are colored based on their sign of DFT binding energies 1NN and 2NN binding energies, i.e., black: 1NN>0 & 1NN>2NN, red: 2NN>0 & 2NN>1NN, and blue: 1NN<0 & 2NN<0.

Figure 9 Lowest formation energies and maximum binding energies in the (a) Cu-X, (b) Mg-X and (c) Si-X systems. The solute cluster configuration pertaining to the maximum binding energy is indicated by separate colors. The elements with maximum binding energy in the 1NN configuration are colored in black while those in 2NN configuration or with negative maximum binding energies are colored in red.

Figure 10 Comparison between solute-solute pair binding and solute-solute-vacancy cluster binding energies: (a) homoatomic (X-X) pairs and X-X-vacancy, (b) Cu-X and Cu-X-vacancy, (c) Mg-X and Mg-X-vacancy, and (d) Si-X and Si-X-vacancy. The diagonal line in each figure represents the same binding energy.

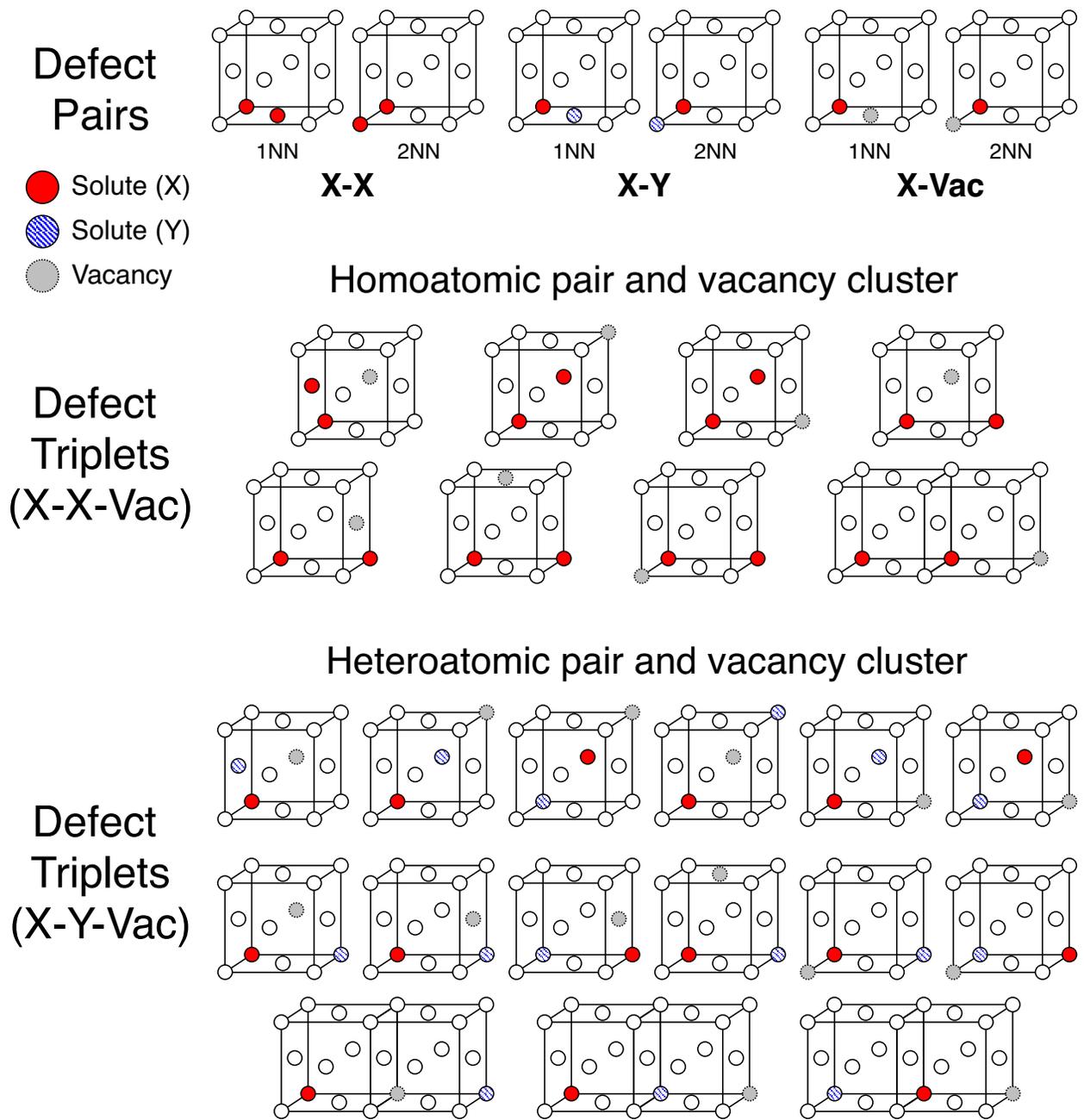

Figure 1 Schematics of atomic arrangements of solute-vacancy, solute-solute pairs and solute-solute-vacancy triplets within the Al supercell. Only the corners of respective 256-atoms fcc Al supercell (4×4×4) are shown.

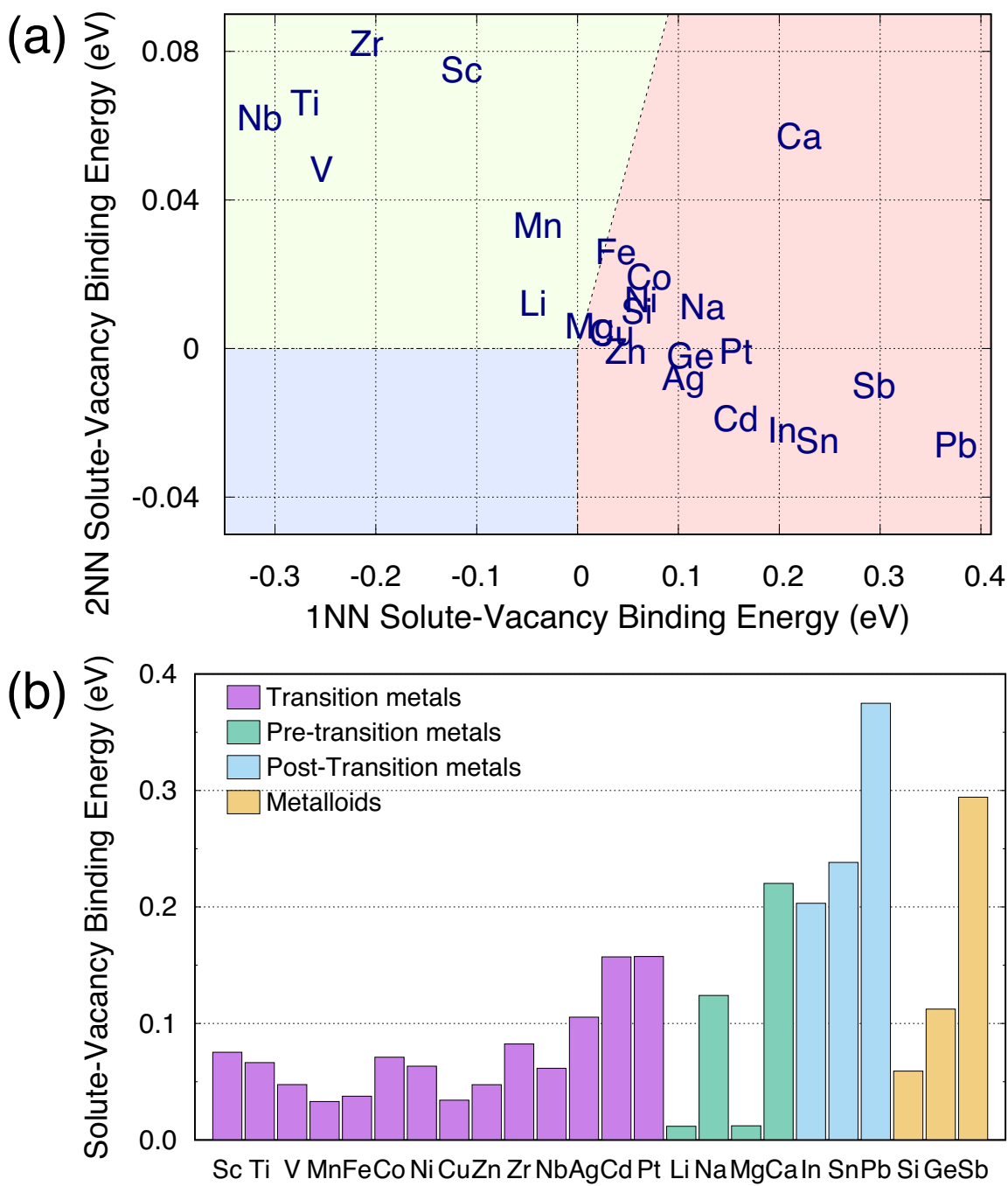

Figure 2 (a) DFT binding energies in 1NN and 2NN for the X-Va pair and (b) maximum values of X-Va binding energies grouped according to the location of X in the periodic table. Within each group, elements are arranged in an increasing order of atomic number. The numerical data is reported in Supplementary Table S1.

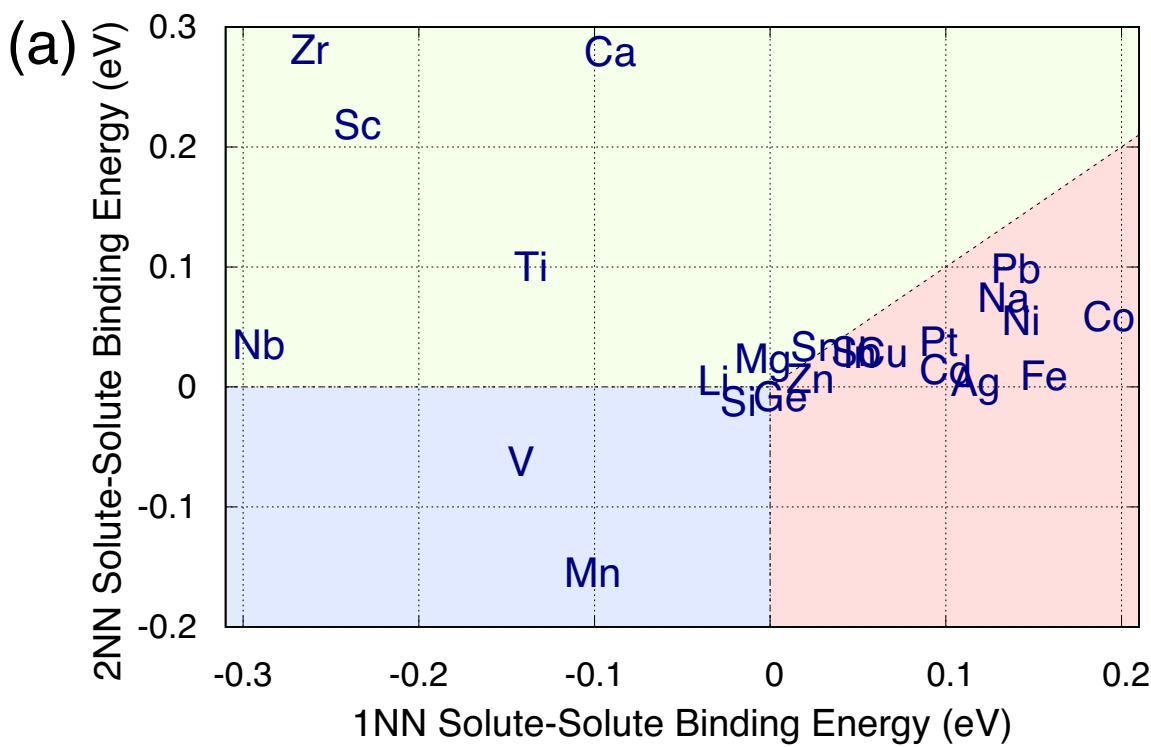

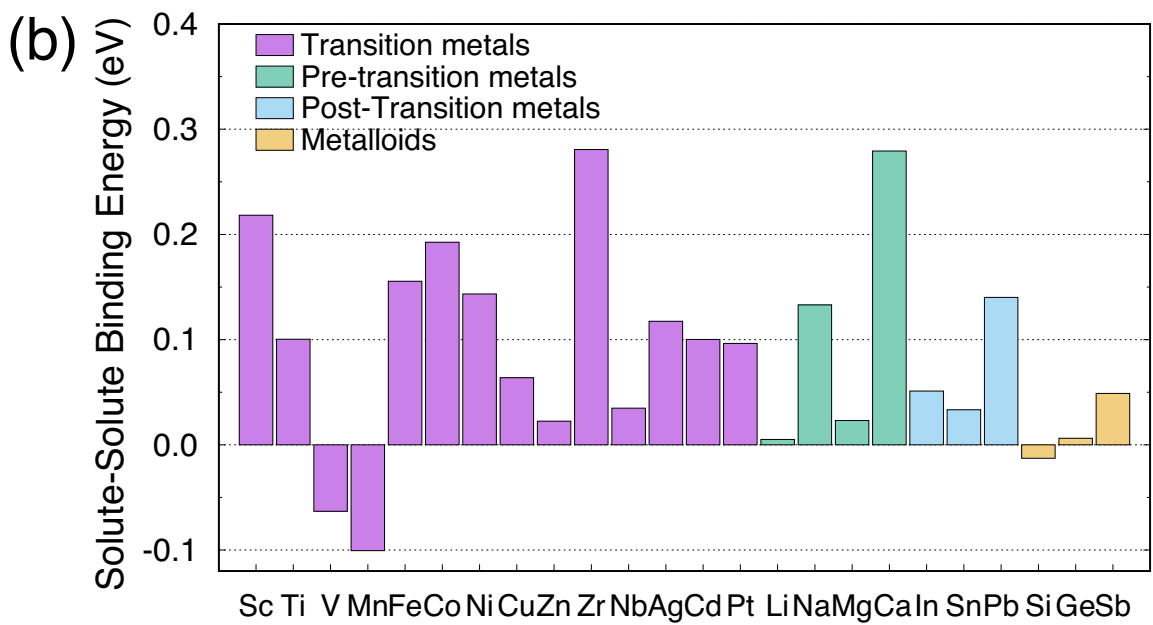

Figure 3 (a) DFT binding energies in 1NN and 2NN for the X-X pair, and (b) Maximum values between the 1NN and 2 NN binding energies according to the location of solute X in the periodic table. Within each group, the elements are arranged in an increasing order of atomic number. The numerical data are reported in Supplementary Table S1.

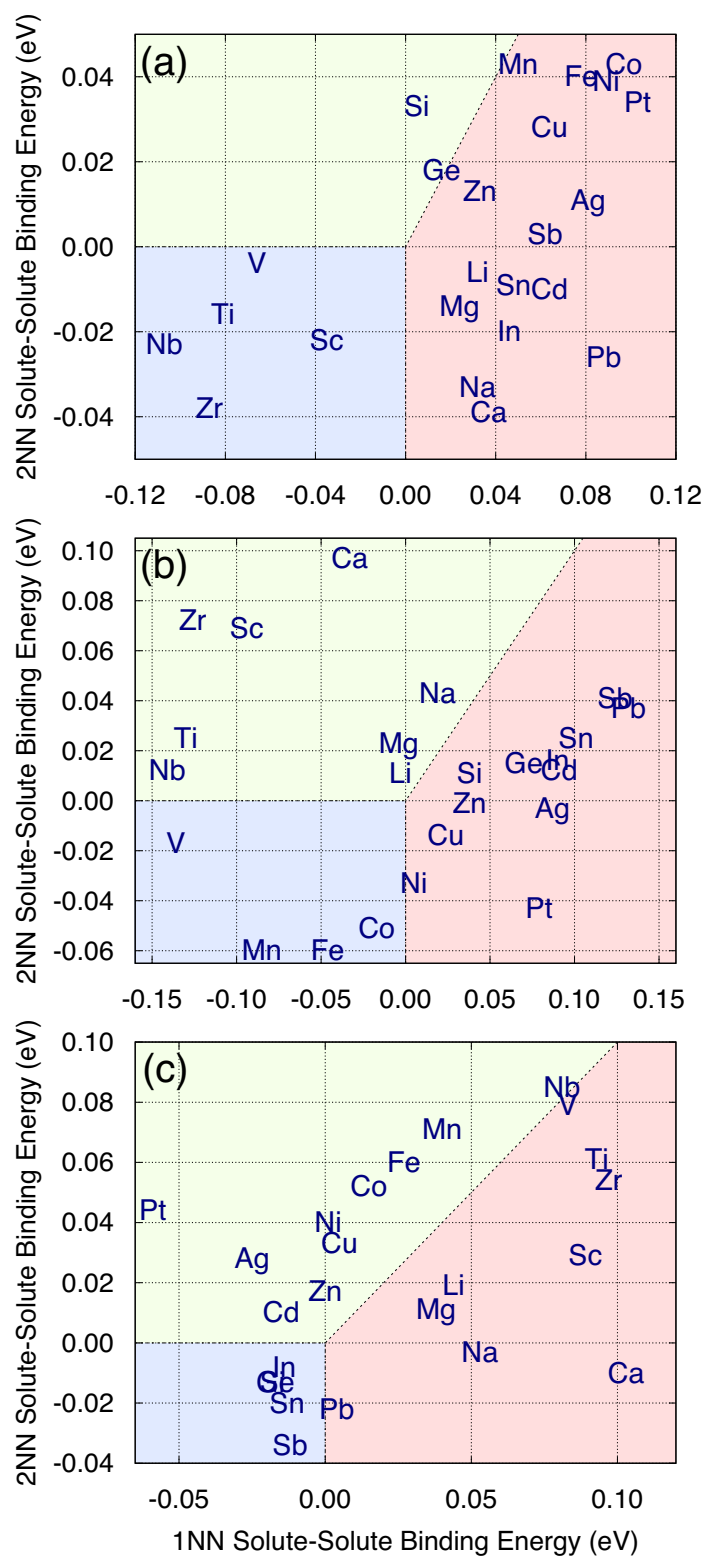

Figure 4 DFT binding energies in 1NN and 2NN for the (a) Cu-X, (b) Mg-X and (c) Si-X pairs. The numerical data are reported in Supplementary Table S1.

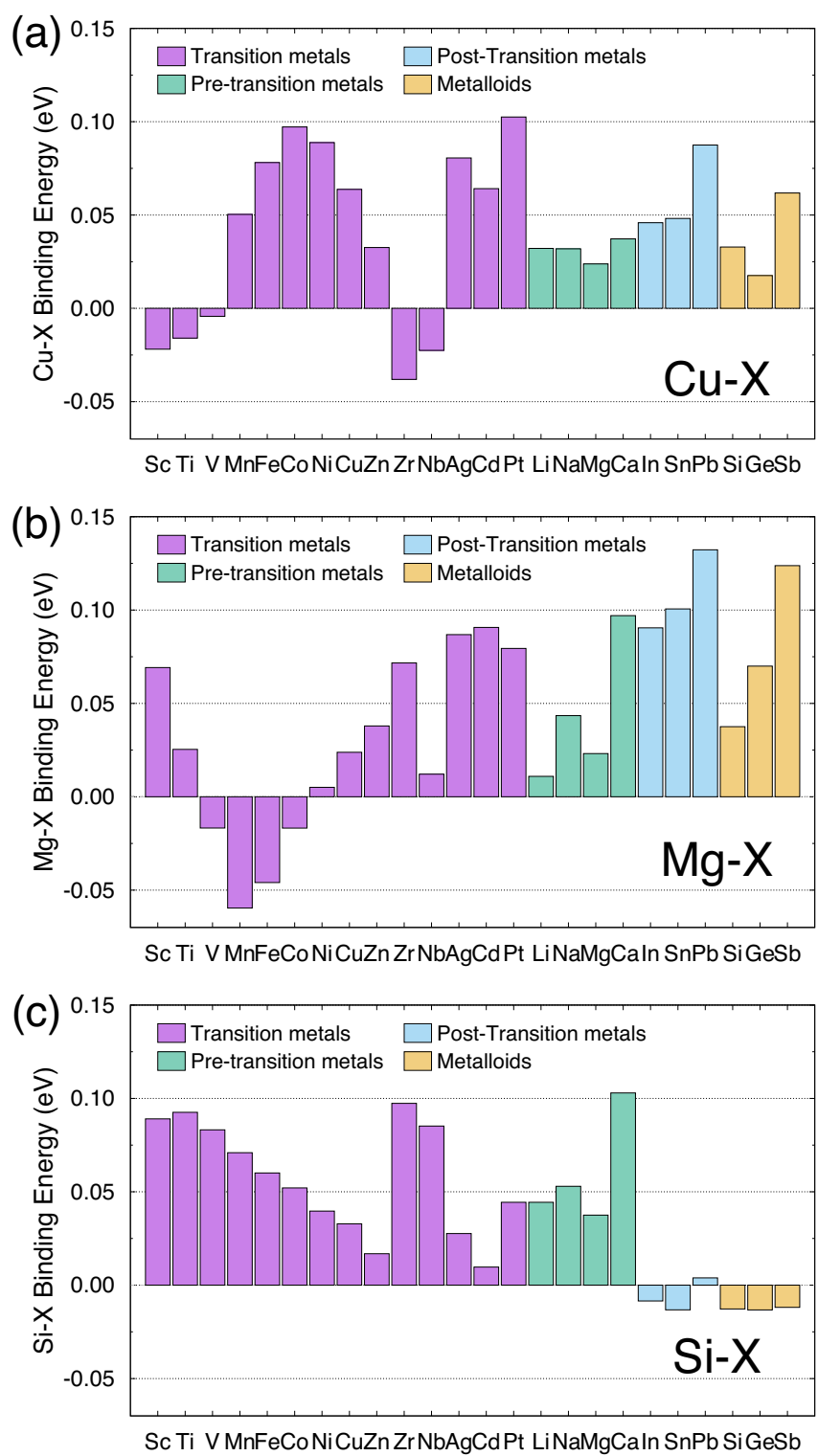

Figure 5 Maximum values between the 1NN and 2 NN binding energies of the Cu-X, Mg-X and Si-X systems grouped according to location of X in the periodical table. The numerical data are reported in Supplementary Table S1.

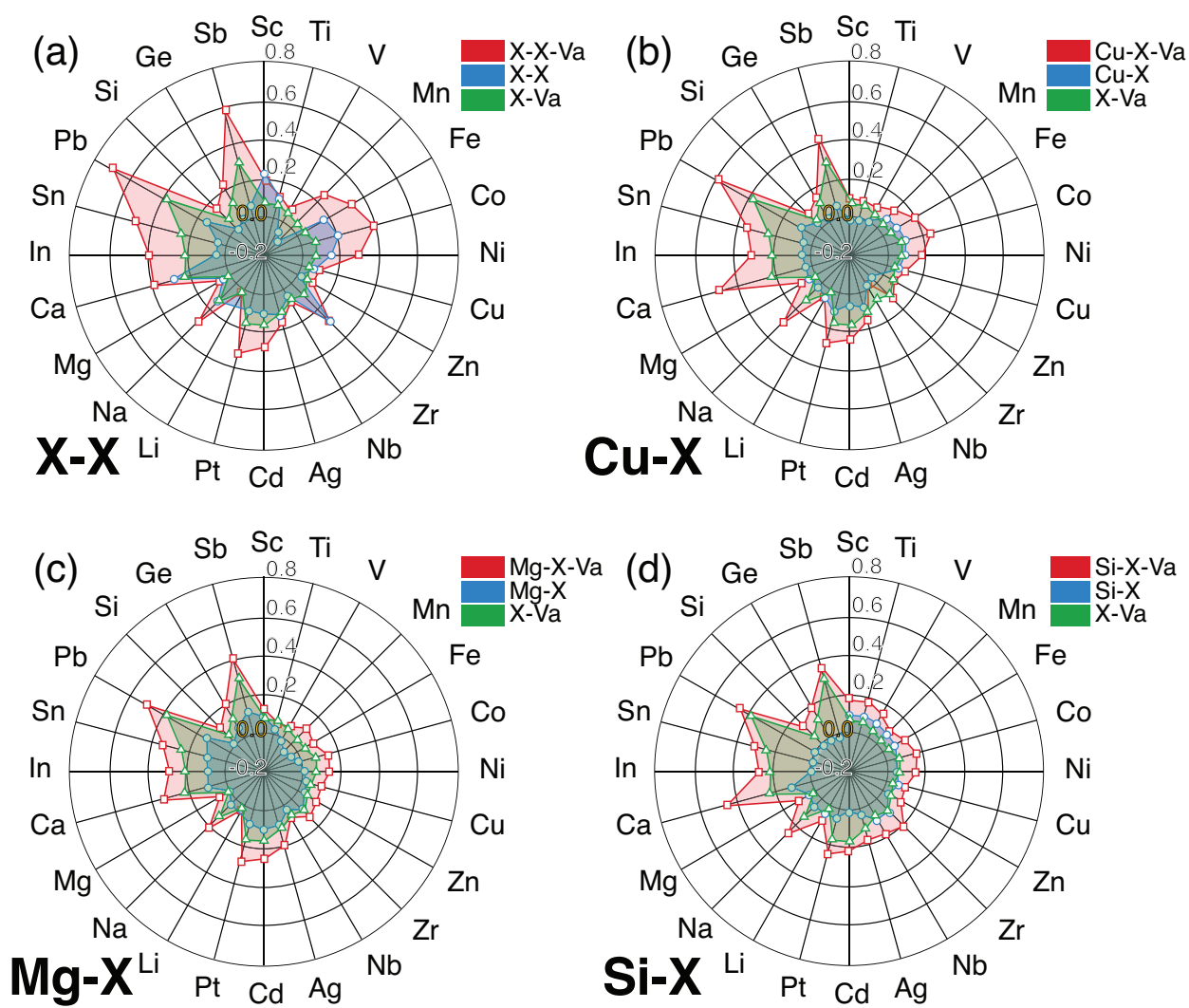

Figure 6 Maximum binding energies (in eV) of various possible grouping in (a) X-X-Va (b) Cu-X-Va, (c) Mg-X-Va and (d) Si-X-Va clusters. The numerical data are reported in Supplementary Table S2.

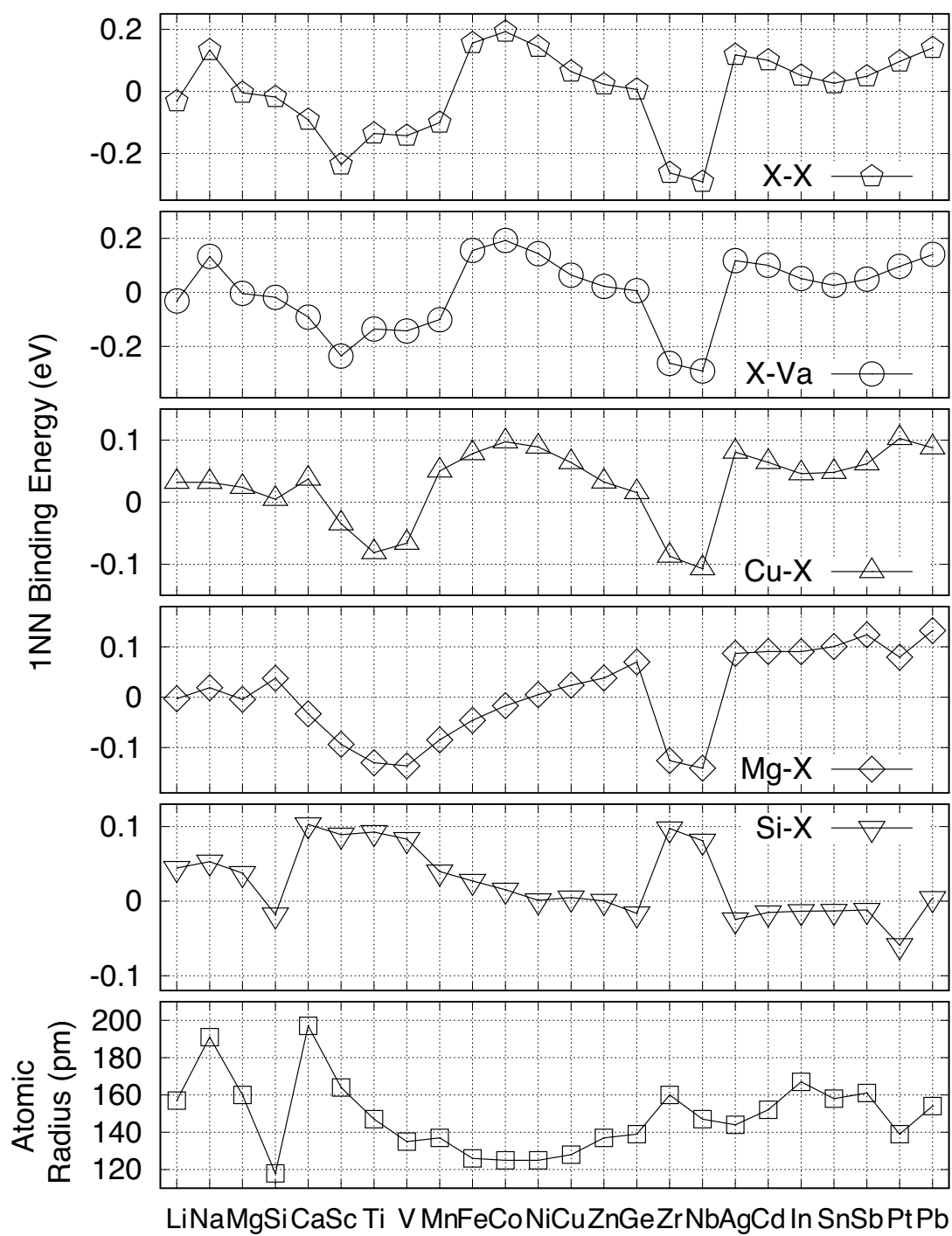

Figure 7 Solute size and DFT 1NN binding energies of the X-X, X-Va, Cu-X, Mg-X, and Si-X clusters.

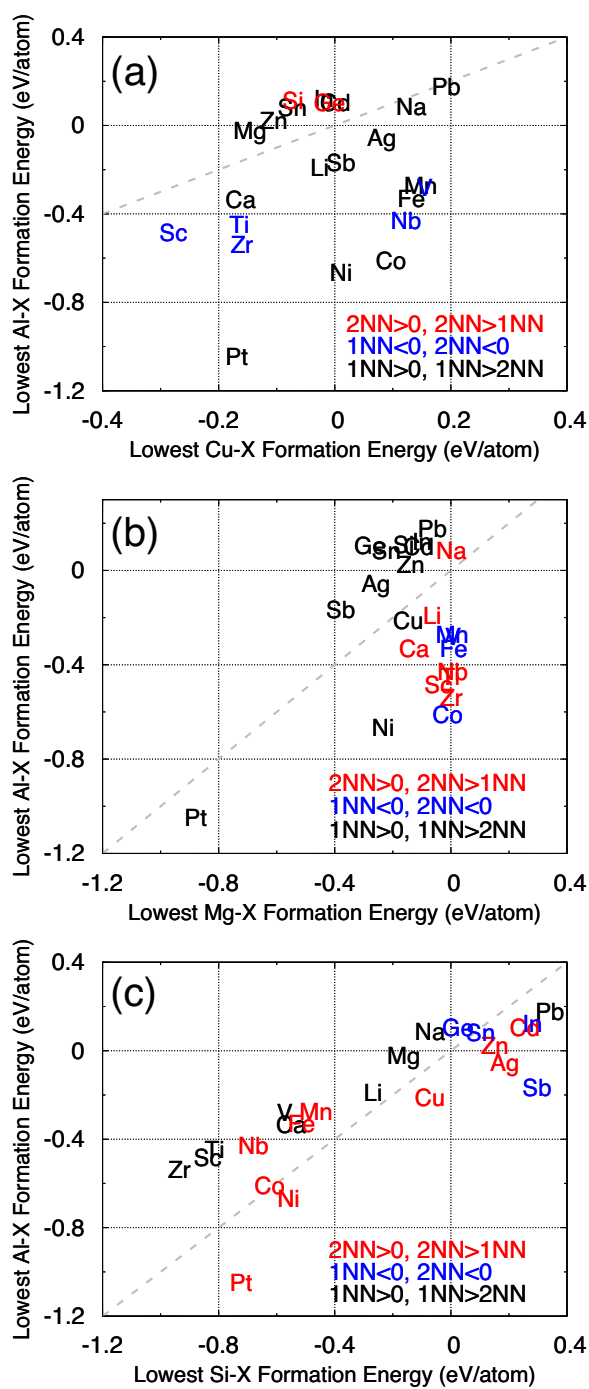

Figure 8 Comparison of the lowest formation energies of the compounds in the (a) Cu-X, (b) Mg-X and (c) Si-X systems with their counterparts in the Al-X system. These data were collected from the open quantum materials database (OQMD) [32]. The black line indicates where the lowest formation enthalpy of the Cu-X, Mg-X or Si-X compounds are equal to that of the Al-X compounds. The elements are colored based on their sign of DFT binding energies 1NN and 2NN binding energies, i.e., black: 1NN>0 & 1NN>2NN, red: 2NN>0 & 2NN>1NN, and blue: 1NN<0 & 2NN<0.

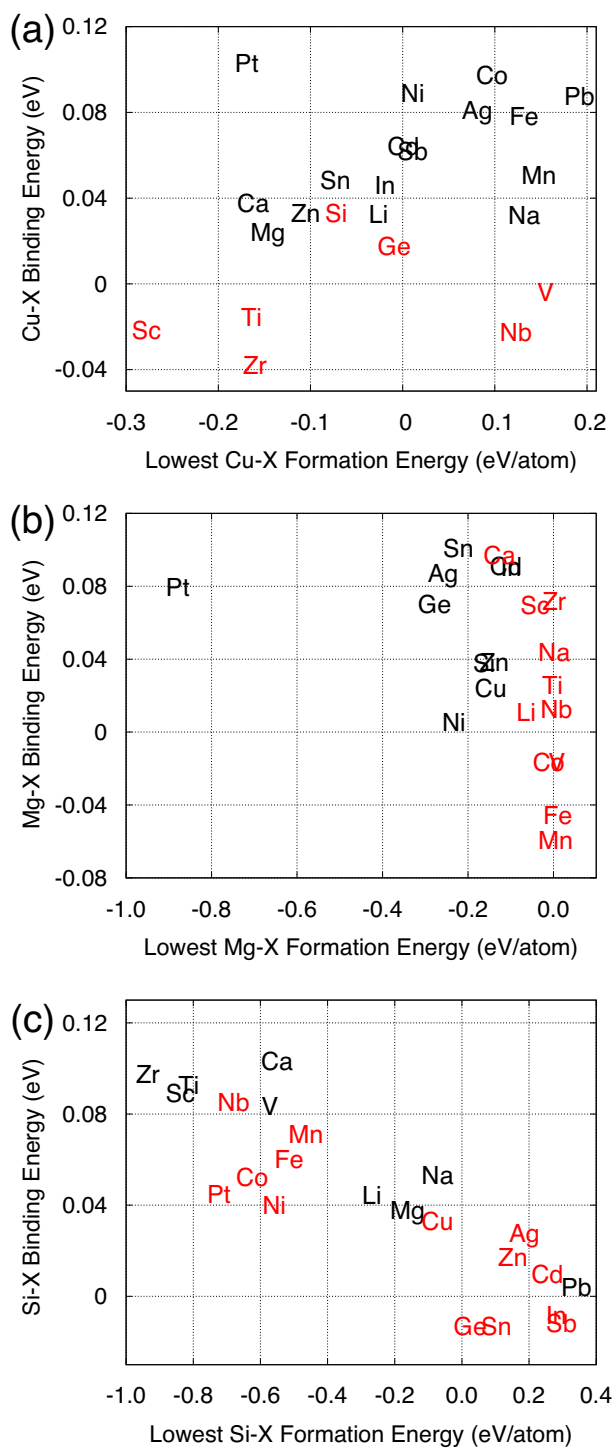

Figure 9 Lowest formation energies and maximum binding energies in the (a) Cu-X, (b) Mg-X and (c) Si-X systems. The solute cluster configuration pertaining to the maximum binding energy is indicated by separate colors. The elements with maximum binding energy in the 1NN configuration are colored in black while those in 2NN configuration or with negative maximum binding energies are colored in red.

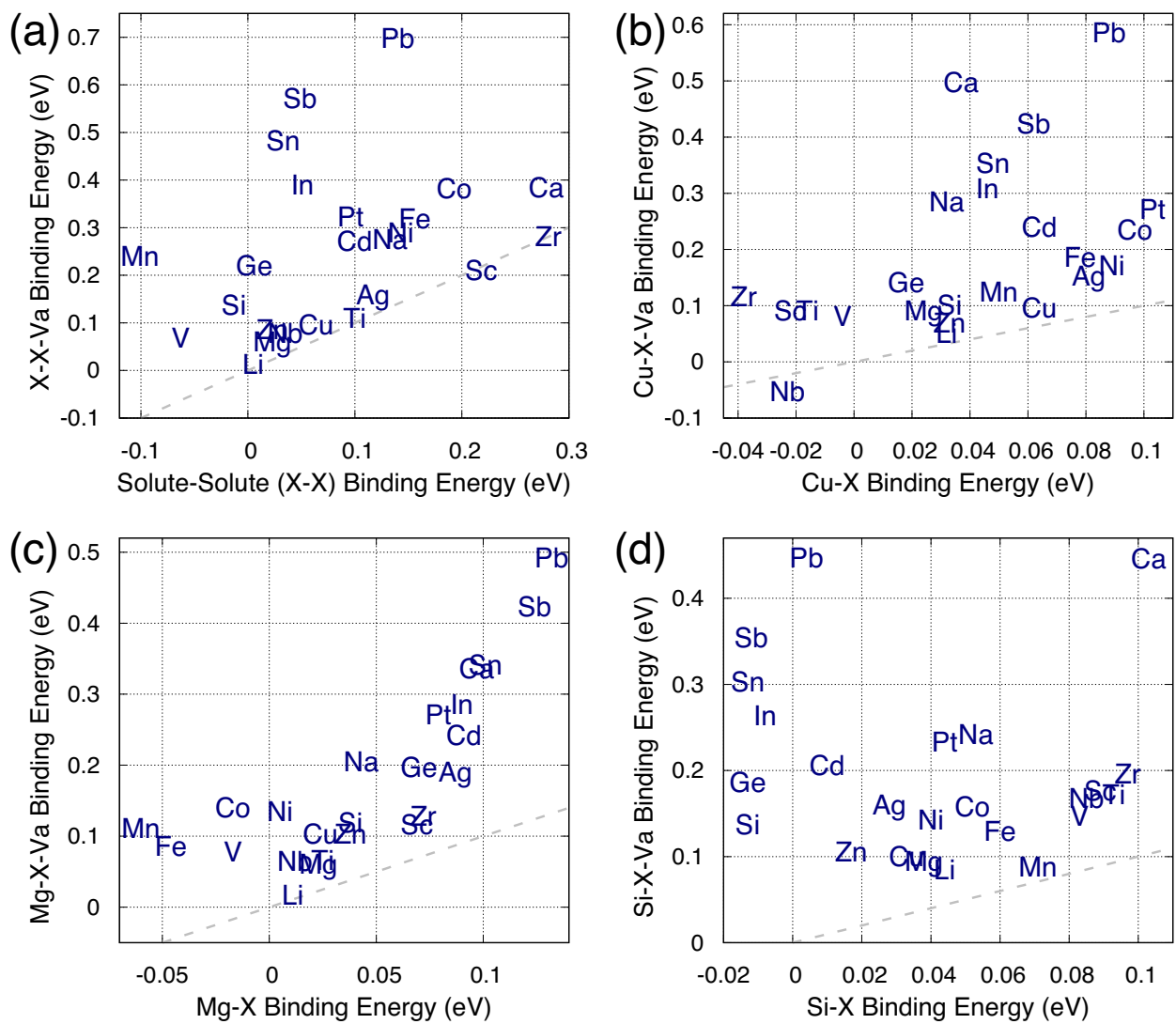

Figure 10 Comparison between solute-solute pair binding and solute-solute-vacancy cluster binding energies: (a) homoatomic (X-X) pairs and X-X-vacancy, (b) Cu-X and Cu-X-vacancy, (c) Mg-X and Mg-X-vacancy, and (d) Si-X and Si-X-vacancy. The diagonal line in each figure represents the same binding energy.

39